\documentclass[11pt]{article}
\usepackage{setspace}
\usepackage{amsfonts,url,epsfig,float}
\usepackage{geometry} 
\usepackage{sectsty} 
\usepackage[normalem]{ulem}
\usepackage{xcolor}
\usepackage{verbatim}
\usepackage{amsmath}

\newcommand{\jessica}[1]{\textcolor{red}{JH: #1}}

\sectionfont{\sffamily\bfseries\upshape\large}
\subsectionfont{\sffamily\bfseries\upshape\normalsize} 
\subsubsectionfont{\sffamily\mdseries\upshape\normalsize}
\makeatletter
\renewcommand\@seccntformat[1]{\csname the#1\endcsname.\quad}
\makeatother\renewcommand{\bibitem}{\vskip 2pt\par\hangindent\parindent\hskip-\parindent}

\makeatletter
\def\@maketitle{%
  \begin{center}%
  \let \footnote \thanks
    {\large \@title \par}%
    {\normalsize
      \begin{tabular}[t]{c}%
        \@author
      \end{tabular}\par}%
    {\small \@date}%
  \end{center}%
}
\makeatother

\newcommand{\btR}{\vspace{-.25in}\begin{quotation}\begin{small}\noindent\begin{verbatim}}

\title{\bf Hypothesizing an effect size by considering individual variation\footnote{We thank the U.S. Office of Naval Research, National Science Foundation, and National Institutes of Health for partial support of this work.}\vspace{.1in}}

\author{Andrew Gelman\footnote{Columbia University}, Amy Krefman\footnote{Northewstern University}, Lauren Kennedy\footnote{University of Adelaide}, and Jessica Hullman$^{\dagger}$ \vspace{.1in}}

\date{7 Apr 2026}

\begin{document}\sloppy
\maketitle

\begin{abstract}
When designing and evaluating an experiment or observational study, it is useful to have a realistic hypothesis regarding the average treatment effect.  We present an approach to conceptualizing this average by first considering a distribution of effects.  We demonstrate with examples in medicine, economics, and psychology.
\end{abstract}

\section{The challenge of hypothesizing an average causal effect}

Hypothesizing effect sizes is important in Bayesian inference for constructing the prior distribution; in frequentist inference for likelihood ratio tests, power calculations, and estimating type 2, type M, and type S errors; in prediction and causal analysis for making inferences about individuals; and for the design of studies, no matter what inference will be done with them.

When designing and evaluating an experiment or observational study, there are several reasons it can be useful to have a realistic hypothesis regarding the average treatment effect.  First, results are interpreted relative to expectations, and when it is time to incorporate a new study into decision making, it is good to have a sense of how its results fit into previous understanding.  Second, for Bayesian analysis a prior distribution is needed, and in many empirical studies the prior is important, because the inferential uncertainty from the data estimate can be comparable to variation in the prior (van Zwet, 2018).  Third, in general the statistical properties of non-Bayesian analyses will depend on the underlying effect size, as for example when evaluating type 2 errors and expected magnitude and sign errors of an estimate (Gelman and Carlin, 2014).  Fourth, in the design stage it is important to have some guess of the underlying effect size in order to assess statistical power and related quantities.

We see problems with existing recommended approaches to hypothesizing effect size:
\begin{itemize}
\item {\em Meta-analysis} or {\em literature review} has the garbage-in-garbage-out problem that the literature will tend to be biased because of selection on statistical significance (Ioannidis, 2008, Button et al., 2013).  It is possible to use Bayesian methods to pool the literature estimates toward zero (van Zwet and Gelman, 2022) but this requires additional assumptions about the distribution of effect size.
\item {\em Expert elicitation} has the problem that experts themselves could well be biased, to the extent that they are relying on the published literature along with their own hopes and beliefs regarding whatever is being studied. We ultimately will recommend a form of expert elicitation, but with a different framing than just asking for an average effect.
\item {\em Conventional designations} of small, moderate, and large effect sizes (Cohen, 1962) can be wildly optimistic in the modern environment of incremental improvements.
\item {\em An estimate from a single study}, even if not subject to selection bias, is in general too noisy to be useful; the problem is that a key reason for hypothesizing effect size is to interpret noisy results (Gelman, 2019).
\end{itemize}
Along with this is a problem in statistical culture, the ``Bayesian cringe,'' referring to a general reluctance to make restrictive assumptions, which results in overestimates of magnitudes of effect sizes and overly optimistic designs.

In previous work, we and others have shown how realistic hypotheses about effect sizes can yield bias corrections, improved inferences (Gelman and Weakliem, 2009, Gelman, Hullman, and Kennedy, 2023), and better designs (Gelman and Carlin, 2014, Wu et al., 2024). Indeed, we have argued that unrealistically weak assumptions, by allowing for implausibly large average effects, have contributed to past and ongoing problems of replication in science.

The present paper addresses the next step, which is how to hypothesize effect sizes in practice.  
The issues discussed here should apply more generally to forming assumptions, hypotheses, or priors about model parameters more generally.  We focus on average causal effects because of their role in policy-making: many studies are designed to estimate the effect of an intervention in order to inform treatment decisions in the world. 

\section{Going beyond existing approaches}\label{problems}

\subsection{Information needed for design analysis}

Canonically prescribed approaches to design analysis solve the problem of determining the sample size needed to detect (at a specified significance level) a hypothesized effect size with some degree of power. 
This requires making several assumptions. 
Typically, these include unbiased measurement and statistical independence of observations (although there are more complicated variants such as measurement error models, spatial correlation, etc.) and assumptions about sample size.

Beyond these structural assumptions, two numbers matter: the average treatment effect and the residual standard deviation of the data.
Perhaps surprisingly, the standard deviation of the data is often the easier of these two things to guess ahead of time.  For binary data with probability near 0.5, the standard deviation of measurement is, from the binomial distribution, approximately 0.5.  For continuous measurements we can often give a reasonable guess of the standard deviation from the scale of measurement or from previous data collection.

Assumptions about the average treatment effect are more challenging.  The treatment effect is unknown; indeed that's typically the reason for doing an experiment or observational studies in the first place.  There are some repeated-experiment settings where a large amount of past data using the same design and population are available, and when that is possible there should be no controversy about using such data to hypothesize possible effect sizes in the next study in the series. However, the present paper concerns the unfortunately common setting where no large clean set of historical data is available.
%an effect size estimate that ignores substantial sources of variability in the populations of interest would be  misleading for assessing practical importance. An effect might account for substantial variation between participants (e.g.h2 ¼ .60) but a negligible proportion between items (e.g. h2 ¼ .04).
%This paper looks somewhat related: https://journals.sagepub.com/doi/full/10.1177/17456916221091565?casa_token=u6Cly5_RZtAAAAAA:-DjDzcm1-y5eSSIHJcgDQTwnve7CC7ySlN2pnSWJmTRvO2NgV-BzcGvbaNyRrfrVpsTpeIjkwNg
%They are arguing that to determine whether an effect is or is not important requires explicitly stating the mechanisms that can amplify the importance of an observed effect size and the mechanisms that counteract it, in addition to any other relevant considerations that might influence how the effect generalizes. 
Researchers use heuristics of varying sophistication for hypothesizing the treatment effect in such settings. These can fail in practice, often by grossly overestimating statistical power, which has serious consequences for design, inference, and decision making.

For example, a common piece of guidance is to look to estimates of the treatment effect in the prior literature and use these in pre-experiment calculations (e.g., Myors, Murphy, and Wolach, 2014).
But research studies can produce highly varying estimates of the same treatment effect even when some aspects of the experiment are held constant (Almatouq et al., 2023), as a result of low statistical power to detect effects, incentives to make bold claims (Scargle, 2000), and researcher degrees of freedom (Simmons, Nelson, and Simonsohn, 2011, Gelman and Loken, 2013).
Combining many prior studies in a meta-analysis does not remove these selection biases, as explained by Szászi et al.\ (2023) in the context of a high-profile meta-analysis of nudge interventions.

\subsection{The problem with using pilot studies to estimate design parameters}

When directly relevant estimates are not available from the literature, researchers sometimes use estimates of effect size and variation from pilot experiments to determine the needed sample size. However, pilot studies are usually run on substantially smaller samples than the main experiment, making these estimates so noisy as to be essentially useless for this purpose (Gelman, 2019).

Consider, for example, a pilot experiment with 200 people, with 100 in the treatment group and 100 in the control. Imagine that the treatment is intended to prevent some event, such as death, among those diagnosed with a disease. 94\% of the treated group survive, compared to 90\% of the control group. The estimated treatment effect is therefore 0.04, with standard error of 0.04. 
Effect sizes ranging from roughly $-0.04$ to 0.12 would be consistent with these data, and this range covers too many design possibilities to be useful on its own, even setting aside potential biases in the pilot study.

To see the problem, imagine a research team whose goal is to power their study to detect the true effect with some probability, say 0.8, and they assume the true effect is captured by the pilot estimate in calculating the sample size that achieves this power. 
We should not trust their estimate of the power to detect the true effect that their study achieves, because it was calculated using a noisy estimate of the effect size.
For example, assume the true effect size is 0.01, corresponding to a true probability of survival of 0.92 for untreated people diagnosed with the disease and 0.93 for treated people. If the researchers assume their pilot estimate of 0.04 reflects the true effect, but the true effect is actually 0.01, then their power estimate will be dramatically overstated, and the sample size they would actually need to reliably detect the true effect is 16 times greater than the one they had calculated.
Once the main study results are in, two scenarios are possible. The researchers may observe a smaller than expected effect which doesn't reach significance, or they may ``get lucky'' and observe a significant result, failing to recognize that because their study was severely underpowered, the observed estimate is biased upward, sometimes by a huge amount (Button et al., 2013, Gelman and Carlin, 2014).

\subsection{From an effect size to a distribution of effect sizes}

There is a principle in mathematics that, if you are stuck on a problem, you can make progress by embedding in a larger problem.  For example, theorems about prime numbers can be proven on the space of ideals, and equations defined on real numbers can be solved on the complex plane.  In statistics, sparse-data problems can be attacked using hierarchical modeling, increasing the dimensionality of the problem and introducing uncertainty but allowing a more stable solution.  For example, the method of meta-analysis can usefully applied to a single study (van Zwet, Wiecek, and Gelman, 2025).

In the present paper we propose embedding the problem of hypothesizing a reasonable effect size into the larger, but paradoxically more tractable, problem of hypothesizing a distribution of effect sizes.

A pitfall of standard approaches to experimental design and sample size calculations is that researchers often want to report multiple effects from a single study, but may fixate on a single effect in design calculations. 
For some applications it is the \emph{smallest} expected effect of interest that should be used in the design analysis.  
Imagine that the researchers designing the study mentioned above decide to also estimate the effect of an interaction between the treatment and a behavioral modification.  Further suppose that this interaction effect is half the size of the main effect of the treatment, with twice the standard error. Then if the researchers want their study to be able to detect the true interaction effect with power 0.8, they should 
increase their sample size by another factor of 16 beyond the 16-fold increase implied by the pilot estimate---roughly 250 times the originally calculated sample size.
Often in practice however researchers power their study for main effects of interest, but report additional effects that would be expected to be smaller than the main effects. 
We should expect these comparisons to be underpowered, potentially severely.

\subsection{Hypothesized effects vs.\ Bayesian priors}
%\jessica{maybe this is better in the set up for why we want to hypothesize a distribution (or can be omitted - inspired by one of Andrew's blog posts)}
One reason that hypothesizing an effect size might seem particularly difficult for many researchers is because effect sizes have generally taken a back seat to emphasis on p-values for rejecting a null hypothesis in the social sciences. 
A hypothesized effect is related to, but is not the same as, a Bayesian prior, another concept that has been frequently misunderstood (e.g., when interpreted as ``subjective''). Considering the reference distribution implied by a Bayesian prior helps illustrate natural sources of heterogeneity that most hypothesized effects fail to capture. 

When modeling a parameter that is replicated many times in the world, such as the concentration of a particular enzyme in the water of North American ponds, we can think about the Bayesian prior we set as our approximation of the population distribution that we imagine drawing from when we take observations of specific ponds. 
When there is only one value of the target parameter, we can often think of the Bayesian prior as our attempt to approximate the true distribution of underlying parameter values considering all possible problems for which our particular model (including this prior) will be fit. While we'll never know what the true prior is, we know it exists, and we can think of any prior that we do use as an approximation to this true distribution of parameter values for the class of problems to which the model at hand will be fit.

The most challenging setting for conceiving of the true prior we are trying to approximate is when there is truly only one value of the parameter that exists. For example, imagine doing an experiment to measure the speed of light in a vacuum. However, even when the inference is singular, it is often still possible to embed the problem into a larger class of exchangeable inference problems. We might consider all settings in which some physical constant is estimated from an experiment. 

We propose to approach the question of what treatment effect to hypothesize similarly to how we define a Bayesian prior in settings without natural replication: by hypothesizing a distribution of effect sizes.

\section{Proposed method: Hypothesize a distribution of effect sizes}

We consider four situations in which one might need to hypothesize an effect size:
\begin{itemize}
\item When designing a study to have a sense of its statistical properties such as bias, variance, power, and type 2, M, and S error probabilities of inferences;
\item When estimating these statistical properties for a study that has already been conducted;
\item When constructing a prior distribution for Bayesian analysis or choosing a range of applicability for a classical method; 
\item When fitting a model with interactions or hierarchical structure so that there is interest in variation in the effect, not just its average.
\end{itemize}

For all these problems, our recommended approach is to construct a model for the \emph{distribution} of effect sizes, even if only the average that is of practical interest.  Paradoxically, forming a denser network of assumptions can be easier than hypothesizing a single number, in the same way that it can be helpful to think about meta-analysis even when drawing inference from a single study (van Zwet, Wiecek, and Gelman, 2025).

\subsection{Hypothesizing a distribution of individual treatment effects}

Our basic strategy uses the decision making heuristic of anchoring and adjustment (Tversky and Kahneman, 1974):  start with a guessed effect size and then make corrections based on assumptions about its variation:
\begin{enumerate}
\item Hypothesize a realistic effect size when the treatment is working as intended.  This starting point can be based on some combination of the scientific literature (if necessary, correcting for publication and reporting bias) and subject-matter reasoning.  We call this the {\em plausible effect size}.
\item Consider a range of effects under real-world conditions.  For example, if the plausible effect size is $X$, the range of effects might be $(0,X)$.
\item Convert this range into a distribution.  For example, a range of $(0,X)$ could be approximated by a normal distribution with center $0.5X$ and scale $0.5X$.  In the absence of additional subject-matter information, this could make sense:  if the ideal effect is positive, there could be rare exceptional cases with even higher effects and occasional negative cases.  A negative effect could happen, for example, if the existence of an effective educational intervention motivates some students to relax and study less.
\item Hypothesize a proportion of pure nulls, people for whom the treatment effect is zero because it is not activated at all.  This could be patients who do not take a pill or who lack the ability to metabolize a certain compound, students who are not paying attention in class, customers who are never exposed to a promotion, etc.  As a default, we might suppose the proportion of pure nulls to be 50\% for direct interventions in medicine and social science, or 90\% in marketing, where interventions are typically ignored.
\end{enumerate}
Putting these steps together, a reasonable first guess of an average treatment effect can be found using  $\widehat{ATE} \approx (1 - p_{\text{null}})\cdot 0.5X$, where $X$ is the plausible effect size. 
This equates to taking the plausible effect size divided by 4 for direct interventions, and divided by 20 (or even slightly more) for indirect interventions.

\begin{comment}
\url{https://onlinelibrary-wiley-com.turing.library.northwestern.edu/doi/epdf/10.1002/sim.694}
\url{https://www.sciencedirect.com/science/article/pii/S0895435609001759}
\url{https://arxiv.org/pdf/2112.01380.pdf}
\end{comment}

This approach encourages researchers to challenge their initial expectations about treatment effects, as is often recommended to elicit experts' knowledge on a quantity in use cases like prior specification in Bayesian modeling (Mikkola et al., 2023) or expert-informed decision making (Garthwaite et al., 2005; O'Hagan et al., 2006).
A takeaway from this literature is that people often struggle to express their knowledge in terms of probabilities, requiring carefully structured elicitation protocols to avoid biasing responses. 
For example, overconfidence manifests as elicited 95\% probability intervals that contain the corresponding true values much less than 95\% of the time.
Asking the person to think about the full range of possibilities for the quantity to be elicited can help counter overconfidence.
Additionally, by focusing attention on heterogeneity at the level of individual units, our approach encourages the researcher to consider their expectations about possible moderators of an effect from the bottom up, helping curb availability bias as a result of failures to consider relevant background knowledge (O'Hagan, 2019).  

%As a result, when hypothesizing and interpreting effects, it is easy to overlook differences in study purposes. 
%Things get worse when the estimated effect size of a study is used to inform the hypothesized effect size in subsequent studies with different purposes.

\subsection{Formative interviews with researchers using the approach}
We conducted formative study of variations of the protocol with three researchers associated with the medical and engineering schools at Northwestern University. We targeted researchers who had a research question related to the treatment effect of an intervention that could be studied in a simple treatment vs control research design (whether between- or within-subjects) for which they planned to conduct a study. Researchers were recruited through our network but were not aware of the specific goals of the project. 

We first asked interviewees to discuss the study they were considering, describing the population individuals will be recruited from, estimated sample size, planned treatment or intervention, the control or comparator, the outcome and how it would be measured, and their analysis plan. Once this contextual information had been laid out, we asked for an estimate of the average treatment effect after the study concludes \textit{if the treatment works as intended}. We refer to this as $\hat{ATE}_{\text{pre}}$.

We next asked them to consider what treatment effects might occur at the individual level, even if these could not be observed in practice. We prompted them to consider the people they might sample with the largest and smallest treatment effects, including their characteristics, the estimated treatment effect they expected for that type, their uncertainty in their estimate, and the proportion of the sample they would expect to have that effect or more extreme. 

We then asked the interviewees to consider the rest of the people in their study. In the first interview, with a researcher studying online misinformation interventions, we used an online distribution builder tool (Chang et al., 2024; Oakley, 2024), where the participant assigns some number of balls (e.g., $n=20$ total) to each of $k$ bins. %However, after the first session we conducted with a researcher studying online misinformation interventions, we became concerned that the distribution builder encouraged the participant to submit beliefs that conformed with distributional expectations  (e.g., that the effect distribution must be normally distributed) rather than considering  different types of individuals and their effects. 
%We therefore did not use the distribution builder for the second and third researchers. Instead, after eliciting the smallest and largest plausible individual effects, 
In the second and third interviews, we simply asked for the proportion of people who would fall between the smallest treatment effect and the midpoint between the smallest and largest, and again between the midpoint and the largest treatment effect. For each participant, we calculated the implied average treatment effect by taking the proportion-weighted average over the distribution of elicited effects, which we refer to as $\hat{ATE}_{\text{post}}$. We then asked the participants to consider $\hat{ATE}_{\text{post}}$ relative to $\hat{ATE}_{\text{init}}$ (if they were different), and to share their comments on which they felt better aligned with their domain knowledge going into study design.

All three researchers were enthusiastic about the idea of being prompted to think through individual treatment effects, even in one case where their initial estimate did not change. One stated, ``I \dots appreciate this exercise, though, for thinking about what are the actual effects, and will they be different for different people, and \dots maybe that also informs \dots who's included in the study as well.'' Another researcher said, ``It's good to be challenged and think about the differences in the potential participants and \dots why the change might differ, to a different degree, yeah.'' In one case, the mixture of effects produced an effect size the researcher deemed implausible. He stated he may have overestimated the number of people who would see the largest effect size. He also noted that there were people who were left out of the sample, but that they'd been left out by design, emphasizing that hypothesizing an effect size well requires us to be clear about the study's stated purpose, as we discuss further below.

While the researchers seemed easily able to describe the characteristics of the people expected to have the largest and smallest effects, they found it challenging to identify who may fall in the middle or near the average, as well as what proportion of their sample belongs in these buckets. This supports the idea of instead constructing a distribution where the range is given by the largest and smallest estimated effects, and the participant only needs to consider the proportion of nulls.

\section{Example scenarios}

\subsection{Binary potential outcomes in medicine}

Consider a binary outcome. Imagine our treatment is designed to prevent a specific outcome, such as death within one year of receiving it. Our outcome variable is therefore the survival rate.

For the first step, we can imagine three categories of patients:
\begin{itemize}
    \item Those who will live regardless of whether they receive the treatment or not.
    \item Those who will be saved by the treatment.
    \item Those who will die regardless of whether they receive the treatment or not.
\end{itemize}

Because our outcome is binary, we do not need to make any numeric judgments about the upper or lower bound of the treatment effect. We proceed to step 2, in which we consider the frequency with which we might see each of the three types above. 

If our goal is to demonstrate the efficacy of the treatment, we want our study sample to include as many people of type two (save-able) as possible. Say we think its reasonable to expect that we will be able to limit type 1 (live regardless) in our sample to no more than 30\%.  We also think we can mostly avoid type 3 by pre-screening. To be conservative, so we assume at the most we might have 5\% type 3. This leaves 65\% type 2.    

In the final step, we choose a hypothetical sample size, and combine the three types at the rates we've hypothesized to estimate the average treatment effect. 

Imagine we start by assuming the largest possible sample size we can imagine recruiting: 10,000 people. Sampling 10,000 observations with the probabilities we set in step 2, we get a distribution of effects ranging from from 0 to 1, with 65\% of individuals having a treatment effect of 1 and the remaining 35\% having a treatment effect of 0, yielding an average treatment effect of 0.65. This corresponds to a treatment group survival rate of 95\% versus a control group survival rate of 30\%.

Now consider what happens when we cannot screen participants as aggressively, a situation more representative of effectiveness than efficacy. Suppose type 1 individuals (those who would survive regardless) make up 60\% of the sample rather than 30\%, reflecting a broader recruited population in which many participants face lower baseline risk. Suppose further that pre-screening for type 3 individuals (those who will die regardless) is impractical, and they constitute 20\% of the sample. This leaves only 20\% type 2 individuals whose survival the treatment can actually affect. The resulting average treatment effect drops to 0.20---less than a third of the efficacy estimate---with a treatment group survival rate of 80\% versus a control group survival rate of 60\%. This contrast illustrates how the same intervention, with the same mechanism, can appear to have a dramatically smaller effect when studied in a more representative population, and why effect sizes from tightly controlled efficacy trials should not be used uncritically as the basis for hypothesizing effects in broader effectiveness studies.

For a second scenario in medicine, we use an example from Zelner et al.\ (2021) of a doctor designing a trial for an existing drug that he thought could effectively treat high-risk coronavirus patients. He solicited our help to check his sample size calculation that a sample size of $n = 126$ would assure 80\% power under an assumption that the drug increased survival rate by 25 percentage points.  (With 126 people divided evenly split between two groups, the standard error of the difference in proportions is bounded above by $\sqrt{0.5*0.5/63 + 0.5*0.5/63} = 0.089$. To achieve 80\% power requires the value of the effect to be at least 2.8 standard errors from the comparison point of 0, hence, an effect of 0.25 achieves the desired power with $n = 126$.)  When asked how confident he felt about his guess of the effect size, the doctor replied that he thought the effect on these patients would be higher, such that 25 percentage points was a conservative estimate. At the same time, he recognized that the drug might not work.  But when asked what he thought about increasing his sample size so he could detect, for example, a 10 percentage point increase in survival, he replied that this would not be necessary:  he felt confident that if the drug worked, its effect would be large.

It might seem reasonable to suppose that a drug might not be effective but would have a large effect in case of success.  But to stop at this assumption implies a problematic vision of uncertainty.  Suppose, for example, that the survival rate was 30\% among the patients who do not receive this new drug and 55\% among the treatment group.  Here is a hypothetical scenario of what we might expect given 1000 people:
\begin{itemize}
\item 300 would live either way,
\item 450 would die either way,
\item 250 would be saved by the treatment.
\end{itemize}
There are other possibilities consistent with a 25 percentage point average benefit---for example the drug could save 350 people while killing 100---but the point is that once we assume a scenario as we did above, the posited benefit of the drug is not a 25 percentage point benefit for each patient; rather, it's a 100\% benefit for 25\% of the patients.

From that perspective, once we accept the idea that the drug works on some people and not others---or in some comorbidity scenarios and not others---we realize that ``the treatment effect'' in any given study will depend entirely on the patient mix.  There is no underlying number representing the effect of the drug.  Ideally one would like to know what sorts of patients the treatment would help, but in a clinical trial it is enough to show that there is some clear average effect. Once we consider the treatment effect in the context of variation among patients, this can be the first step in a more grounded understanding of effect size. 

\subsection{Numerical-data example in political science}

In Gelman and Margalit (2021), we studied the changes in political attitudes that arise from knowing people in a relevant social group.  For example, what is the effect of knowing a gay person on attitudes toward same-sex marriage, or how much does knowing a recently-unemployed person affect views on unemployment insurance?  We collected two waves of data a year apart on 1700 respondents, asking about penumbra membership and issue attitudes both times.  For each of 14 issues, we fit a regression to the respondents who were not in the relevant penumbra, predicting attitude at time 2 given penumbra membership at time 2, also adjusting for attitude at time 1 and demographic background variables.  Attitudes were measured on a 1--5 scale.  The estimated effects varied by issue and were mostly in the range 0.05 to 0.1 with standard errors of about 0.05.  These estimates were small (no more than a tenth of a point on five-point scale) and just barely large enough to be distinguishable from pure noise.

We were relieved to find a detectable signal in our data but were surprised that the estimated effect was so small.  In retrospect, it would have been a good idea to think more carefully about effect sizes before the study began.  We can do so here.

Start with that 1--5 scale.  On contentious political issues, many respondents will be already at 1 or 5 and will not be moving from those extremes.  What about those whose views are changed by entering the penumbra?  Some people will have their views changed in a positive way by knowing someone in the relevant group, while others will have negative experiences.  Putting all these together:  perhaps a third of the population already has strong negative views on the target issue, a third have strong positive views, and a third are open to be changed.  Of that third, perhaps 50\% will be unaffected by the entering the penumbra, 40\% will be pushed in a positive direction by one point on the scale, and 10\% will be repelled and pushed down one point on the scale.  Putting this together, the average treatment effect would be $\frac{1}{3}(0.5\cdot 0 + 0.4\cdot 1 + 0.1\cdot (-1)) = 0.1$.

This calculation has several arbitrary choices, and we are not claiming that one can accurately hypothesize the effect size before doing the study.  Rather, our point is
that a bit of thought can give us a sense of reasonable effect sizes under different assumptions, and this can be used to design a study as well as to interpret the results of a completed analysis.

\subsection{Anticipating an effect size in the context of the goals of a study}

A typical effect size calculation would require an estimate for the expected effect size. However, when we consider heterogeneous effect sizes, our research purpose is also required. The research purpose will help to determine which effect size to account for. 

For example, Gelman, Hullman, and Kennedy (2023) we present two sets of graphical quartets, inspired by Anscombe (1973), that we argue should be helpful helpful for planning for and understanding causal effects. Here we consider how we might go about designing a study to compare the effect of this presentation to that of a verbal description of types of causal heterogeneity. If we were planning this study, how would the study purpose interact with how we think about effect variance? 

Suppose the effect size is largest for those with less statistical training and is close to zero for people with high existing expertise, as shown in the reversed S-curve in Figure~\ref{fig:example_effect}.

\begin{figure}
\centering
\includegraphics[scale=0.5]{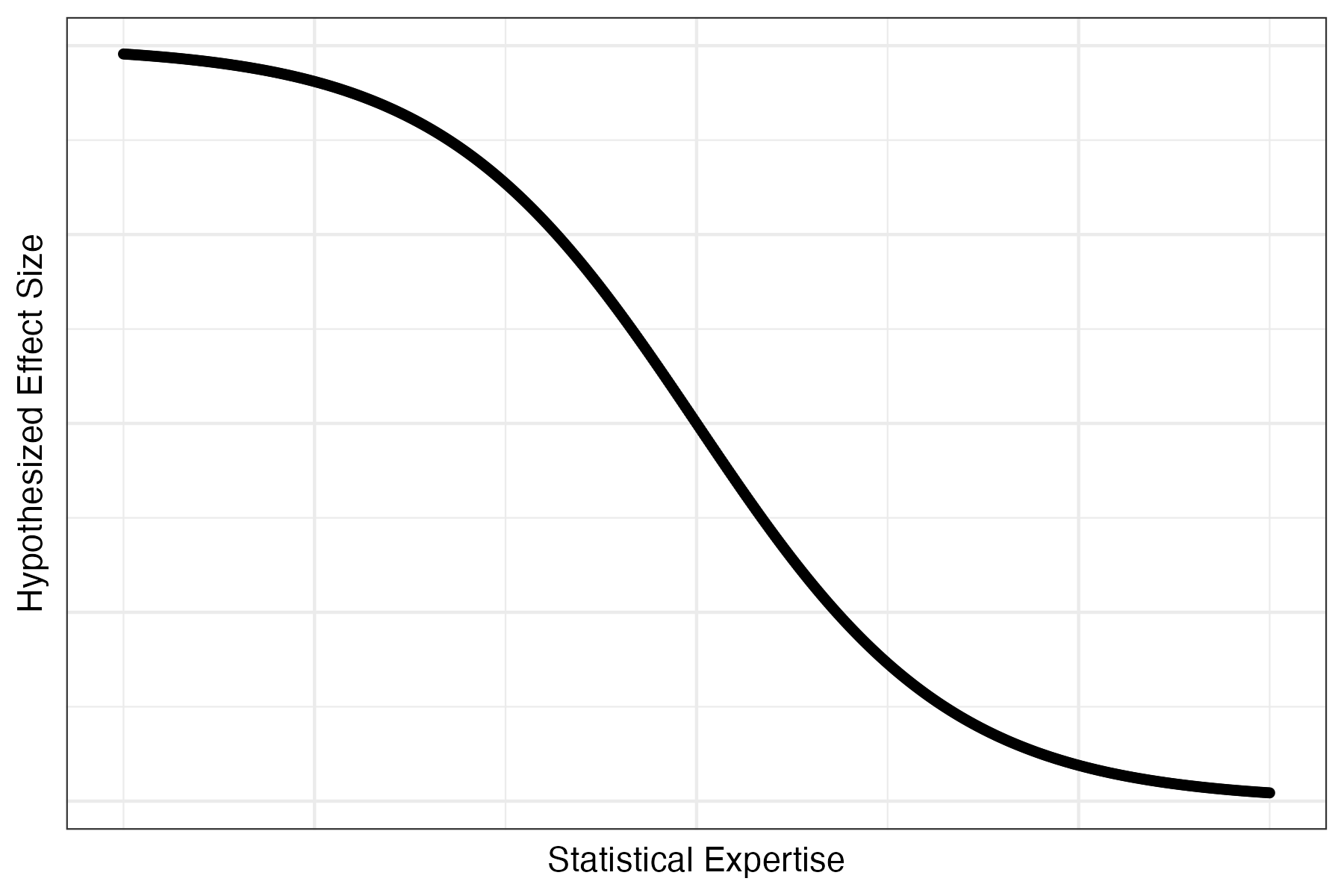}
\caption{Hypothetical relationship between level of statistical expertise and effect of the causal quartet plots. This relationship is not based on previous literature.}
\label{fig:example_effect}
\end{figure}

If the purpose of our study is to detect whether there is ever an an effect of our quartets, we would want to focus our sample on the individuals with less statistical training where the effect size is largest. This would give us the greatest efficiency per participant to observe the effect. 

If the purpose of our study is to estimate the average treatment effect over all people, we would need to identify some method to randomly sample all individuals, or we would need to adjust our sample to the distribution of statistical expertise in the population.
When generalizing beyond the observed sample, it is important to account for variation across people and scenarios and changes over time; this can be done by fitting a model accounting for key pre-treatment variables and then poststratifying to estimate the average treatment effect in the new setting, an idea that has arisen in many fields, including Hotz, Imbens, and Mortimer (2005) in economics, Tipton and Olsen (2018) in education, and Kennedy and Gelman (2021) in psychology.

A third purpose of our study might be to estimate how the treatment effect varies over statistical expertise. With this purpose, if we want to have the same uncertainty for the effect size over all individuals, we would need to over-sample individuals with higher statistical expertise because their expected effect size is smaller. 

All three aims (which is not an exhaustive list, but merely an example) are valid. The first aim would be most efficient in terms of participant hours, and perhaps might be best employed in early stage research: it wouldn't make sense to spend the additional resources to estimated a population average treatment effect if we haven't first confirmed that the effect exists under ideal conditions.  But then some combination of further data collection and assumptions would be necessary to generalize to the population.

\subsection{Downgrading an apparently very large effect}
We can also use the individual treatment effect protocol to interrogate published results that appear too good to be true. 

For example, Gertler et al.\ (2013) performed a randomized evaluation of an early-childhood intervention program, yielding an estimate that the program increased adult earnings by 42\%.  This sounds a bit too good be true, even more so when considering it as an {\em average} effect, given that the actual effect must surely vary a lot by person, considering the tortuous path from an intervention at age 4 to earnings at age 24.  A realistic scenario might be some mix of effects that are often negligible and can follow a wide range when positive, and an effect that is larger in some intermediate zone.  In any of these cases, we would argue that an average effect of 42\% is hard to believe, given that it would reflect some combination of many effects near zero and some increases in earnings of 100\% or more.

The implication of this reasoning is that the claimed effect is likely to be a wild overestimate---a point that we earlier made on inferential grounds (Gelman, 2018) but without reference to varying effects.  Combining a realistic sense of the average effect size with an understanding of selection on statistical significance makes it clear that the study had low power and will yield a positively biased estimate (Button et al., 2013).  The framework of nonconstant treatment effects gives us another reason to be skeptical about the claims made for this particular class of interventions.

%\subsection{Recognizing that an apparently large effect can be explained as an artifact of noise}

As a second example, Beall and Tracy (2013) performed two small surveys and found that women were three times as likely to wear red or pink during certain days of their monthly cycle.  This result achieved conventional levels of statistical significance, but this could easily be explained by uncontrolled researcher degrees of freedom; see Simmons, Nelson, and Simonsohn (2011) for a general discussion of this issue and Gelman (2013) in the context of this particular study.

Rather than statistical significance, we consider the reported effect size, which is implausible on its face and even more outlandish when considered as an average effect, once we reflect that the effect will be zero for many people, for example, those who never wear red clothing or those whose clothing choices are restricted because of work.  Even if we think it's reasonable to expect a factor-of-3 effect for some women in the study, the average effect including those with no effect would have to be much lower, indeed in this case lower than the uncertainty in the estimated effect.
%\jessica{Since the section is called recognizing that an effect can be an artifact of noise, it would be nice to say approximately what the SE was. When I looked back at the paper to calculate it, I found I get a different odds ratio than they report for Sample A (3.85) from the numbers they give... (17 out of 100 women wearing pink or red, 76\% of which are high fertility... should be OR of 4.04 unless I'm somehow mistaken?) Anyway, 95\% CI on this OR is 1.21 - 13.4}. 
%% AG:  The above details seemed like a distraction to me, so instead I added a bit to the first paragraph of this example to clarify that the reported result was statistically significant.
This implies that the published result, despite its apparent statistical significance, could be explained by a combination of chance and unintentional selection bias.  Indeed, followup studies by these authors and others did not replicate the finding; see, for example, Hone and McCullough (2020).

Beyond all this, if time of the month does influence clothing choices, we would expect this effect to vary greatly across people and scenarios.  There is no theoretical reason to expect a common direction, hence a mix of positive and negative effects seems likely, to the extent there are large effects at all.  Such variation makes it even more difficult to estimate an average treatment effect, as well as suggesting that any realistic average would be close to zero.

In earlier writings, we have used the day-of-cycle and clothing study as an example of the perils of naive interpretation of statistics.  Here, thinking about varying effects helps us understand why estimating an average effect for this problem is not well motivated:  the difficulty is not just the lack of successful replication but rather the conceptual framework under which the effect is characterized by a single number or even a single direction.

%\subsection{Anticipating the decline effect: Treatments that are less effective in real life}

\section{Discussion}\label{discussion}
% Living your life in the context of this paper\
%heterogeneity is your friend
A premise of our argument is that treatment effects should generally be assumed to vary across individuals and contexts rather than being characterized by a single underlying quantity. Taking this premise seriously shifts conventions of study design and interpretation. Instead of beginning with a focused hypothesis about a single average treatment effect and only later asking whether effects vary, researchers would start by considering the distribution of plausible individual effects and the composition of the sample being studied. The hypothesized average effect would then emerge as a consequence of assumptions about heterogeneity, study population, and research purpose. In this sense, our proposal encourages an inversion of the typical workflow: rather than treating effect heterogeneity as a secondary complication introduced after detecting an average effect, we treat heterogeneity as the starting point for thinking about design, interpretation, and generalization.

Here we reflect further on the relationship between the hypothesized effect size and research purpose and stage of research. We consider how existing evidence can be used to inform estimates indirectly, and comment on the link between our approach and the replication crisis.

\subsection{Meta-lesson on the relationship between study design, purpose and hypothesized effect}

A takeaway that our perspective highlights relative to conventional approaches to estimating treatment effects is how much our estimates should vary with the composition of the sample that is recruited, and consequently the purpose of our study. 

Our experiences advising researchers in planning studies suggest to us that many people's estimates of treatment effects are not properly conditioned on the stage of research that they are actually in. 
We should expect the relationship between the studied sample and the target population to vary depending on the stage of research. Initial experiments typically draw on convenience samples, followed by larger replications on convenience samples, followed by attempts to estimate effectiveness on representative samples (Bryan, Tipton, and Yeager, 2021). Similar to laboratory experiments in the biological or physical sciences, where the goal is to examine the theory under (unrepresentative) circumstances ``in which sensitive, calibrated measuring instruments are used in an environment carefully freed of forces that might intrude,'' a researcher who is convinced of the value of their intervention will generally see it as their duty to produce the most striking evidence of the effect that they can.
However, researchers often overlook the difference between efficacy (the ability to produce a desired or intended result) and general effectiveness (the degree to which a treatment is successful once deployed widely) when they consider estimating treatment effects (Singal, Higgins, and Waljee, 2014).
The problem arises when the researcher's process of  designing a test for a new idea, which may be based on ``implicit reasoning that they have not yet articulated even in their own minds'' (Bryan, Tipton, and Yeager, 2021) is overlooked, and the estimated effect interpreted as a population average treatment effect rather than a proof of concept or severe test.

For example, when designing a medical trial, and in early stage research in other domains, the first goal is to maximize statistical power.  We say this not cynically but out of a realistic understanding that success---in the form of statistical significance at the conventional level---can be necessary for approval of a drug or procedure, so if you believe your idea is a good one, you want to design your experiment to have a high chance of demonstrating that it works.

Methods of increasing statistical power in an experiment include:  (1) increasing the sample size, (2) improving the accuracy of measurements, (3) including additional pre-treatment predictors, (4) performing within-person comparisons, and (5) increasing the magnitude of the average treatment effect.  Assuming the first four of these steps have been done to the extent possible, one way to achieve the fifth step is to restrict the participants of the study to those for whom the expected effect is as large as possible. %\jessica{aside: there should be name for this ... I find myself bringing it up a lot in conversation}  To the extent that the treatment effect varies in a predictable way, as in Figure \ref{causal_quartet_2}, this can be done.

There is nothing wrong with performing this sort of restriction when designing a study---indeed, it makes a lot of sense in any experiment to focus on scenarios where the signal is highest---and the result should be a higher average treatment effect among participants in the experiment.  When generalizing to a larger population, however, some modeling is necessary conditional on any information used in participant selection.  Thinking about variation in treatment effects makes this clear:  the average effect is not a general parameter; it depends on who is being averaged over.

\subsection{Drawing on prior estimates}
In repeated-experiment settings where a large amount of past data using the same design and population are available, one should certainly use such data to hypothesize possible effect sizes in the next study in the series. On the other hand, even when there is truly only one value that could exist for the parameter we want to estimate, it can help to think about a larger class of exchangeable problems. Most cases will fall somewhere between these extremes, leaving the researcher to decide how much to consider the existing literature. 

We should expect a tradeoff between the direct relevance of the prior work to the focus of the study at hand and the robustness of the prior estimate. Given the garbage-in-garbage-out problem, researchers should distinguish between carefully estimated empirical benchmarks over more closely related interventions that may be biased.
One can look for known empirical benchmarks to serve as a reference whenever there is a strong precedent for studying the effects of interventions in the domain of interest. For the reasons discussed in this paper, such benchmarks will not in general provide direct estimates of average effects, but they can still serve as comparison points to ensure that postulated estimates are in the right ballpark. Ideally these are effects for which there are incentives to attain a robust estimate.

%normative expectations for change in the absence of intervention
For example, when studying educational interventions applied in a classroom setting, one might look to the expected change of test scores after one year of academic growth and general life maturation (Hill et al., 2008). Such estimates are in demand as states use to evaluate the impact of budgeting decisions. 
Meta-analytic estimates of the heterogeneity to be expected in the effect of an educational intervention as a result of differences in instructor, discipline, or grade levels (Taylor et al., 2018) can also be informative. When studying effects on life expectancy, well-established factors that reduce lifespan like heavy smoking (8 years; Streppel et al., 2007) can be used as a rough upper bound.  There is an informal principle in epidemiology that the interaction between X and smoking is larger than the main effect of X, for almost any X.

Whenever reference effects from the literature are used to inform effect size estimates, care should be taken in anticipating the impact of specific study details. 
For example, standardized effect size measures like $d$ are intended to make it possible to compare effect sizes despite different study designs and operationalization of measures. However, 
%Standardized effect size measures are sensitive to the choice of the measurement of interest and the specific study design (e.g., use of repeated measures), because these factors influence the sample variance. 
both standardized and unstandardized estimates of effect size will be influenced by how the researcher samples the units of analysis (e.g., people) and defines the conditions for study. If the size of an effect varies systematically with features of the units, then how we sample the population of interest matters. If the size of an effect varies systematically with aspects of the experimental conditions, such as the stimuli, then the effect size will be conditional on the range of stimuli space sampled. 

Even in the unusual case where there is a fixed effect across units, any standardized measure of effect size will vary depending on the subset of the population of interest that is sampled: sampling from a truncated distribution will reduce the standard deviation and increase the effect size, while sampling from the tails will do the opposite. 

The impacts of these kinds of ``range restrictions'' (Baguley, 2009) may seem obvious in hindsight, but the importance of considering where you sit in the distribution of the effect across the population or conditions of interest is rarely emphasized in advice on how to do design calculations. 

\subsection{The replication crisis}

The ideas of this paper have several points of connection to the replication crisis in science:
\begin{itemize}
\item Most immediately, in a world of varying effects, there is no particular interest in testing a null hypothesis of exactly zero effect, and we should be able to move away from the idea that a ``statistical significant'' finding represents something that should replicate; see, e.g., McShane et al.\ (2017).
\item As illustrated in some of the examples above, when we think about how an effect can vary, we often lower our expectations of its average effect, which in turn can make us aware of problems of low power. % For example, if a study is designed under the naive expectation of an effect size of 0.5, but then on reflection we think that an average effect of 0.1 is more plausible, then the study would require 25 times the sample size (or measurements that are 5 times as accurate) in order to maintain the desired power.
\item Moving away from the framing of ``the'' treatment effect helps us think about variation.  Instead of classifying a new study as an exact replication (with the implication that the effect should be the same as in the original study) or a conceptual replication (with the hope that the effect should have the same sign), we can think of the first study and the replication as representing two different collections of participants and situations. Indeed, heterogeneous treatment effects may suffice to explain many instances of observed irreplicability (Stanley, Carter, and Doucouliagos, 2018, Linden, 2019, McShane et al., 2019, Kenny and Judd, 2019, Linden, 2019, and Linden and Hönekopp 2021, as cited in Bryan, Tipton, and Yeager, 2021, and Tipton et al., 2023).

\end{itemize}
As we have argued elsewhere (Gelman, 2015), ``once we accept that treatment effects vary, we move away from the goal of establishing a general scientific truth from a small experiment, and we move toward modeling variation (what Rubin, 1989, calls response surfaces in meta-analysis), situation-dependent traits (as discussed in psychology by Mischel, 1968), and dynamic relations (Emirbayer, 1997). We move away from is-it-there-or-is-it-not-there to a more helpful, contextually informed perspective.''

%For example, consider a hypothetical experiment yielding an estimated treatment effect of 0.003 with standard error 0.001, in a setting in which an effect size of 0.1 would be large.  One might first want to dismiss the result as ``statistically significant but not practically significant''---but there are various scenarios under which even a small effect would be notable if its sign is well identified.  In an A/B testing setting in a large company, even an effect of 0.003 could represent many dollars, and in social science we might be interested in the direction of an effect (for example, knowing whether people under stress performed better or worse on a certain task) more than its magnitude.  In such an example, our concern would be that, even if the effect is accurately estimated at 0.003 for this particular experiment, it could easily differ for a new group of people in a different environment.  Perhaps the effect would be $-0.004$ tomorrow, $+0.001$ the next day, and $-0.002$ the day after that.  The relevant comparison is not to the standard error---although that does give us a baseline level of uncertainty---but to changes among people, across scenarios, and over time.  Some of this can be learned from data, other aspects of this variation need to be assumed---but there is generally no good reason to assume that the variation in the treatment effect is zero.

A slightly different argument is that in some applications we really only care about the existence and sign of an effect, not its magnitude:  knowing that an intervention works, even a small amount, could give insight and be relevant for future developments.  But the same problem arises here as before:  there is not necessarily any good reason to believe that a small positive effect in one study will apply elsewhere.  It is not clear how to interpret an average treatment effect, even in a clean randomized experiment, without considering how the effect could vary across people and scenarios and over time.

\begin{comment}
\subsection{Recommendations for design and analysis}

Looking forward, how should this affect applied research?

To start, with smaller average effect sizes than previously imagined, better designs are needed:  more accurate measurements, better pre-treatment predictors, larger sample size, and within-unit comparisons.

When moving to analysis, interactions are important but hard to estimate with precision.  So when we do include interactions in our model, we should estimate them using regularization and not demand that they attain statistical significance or any other threshold representing near-certainty.

Conversely, when we fit simple models without interactions, we should not expect that the local average treatment effects being estimated to immediately generalize.  Instead, when generalizing we should allow for both predictable and unpredictable variation in effects, even if in doing so we need to hypothesize scales of variation without direct evidence from the data at hand.

\end{comment}

\section*{References}

\noindent

%\bibitem Almaatouq, A., Griffiths, T. L., Suchow, J. W., Whiting, M. E., Evans, J., and Watts, D. J. (2022). Beyond playing 20 questions with nature: Integrative experiment design in the social and behavioral sciences. {\em Behavioral and Brain Sciences}, 1--55.

\bibitem Anscombe, F. J. (1973).  Graphs in statistical analysis. {\em American Statistician} {\bf 27}, 17--21.

%\bibitem Anvari, F., Kievit, R., Lakens, D., Pennington, C. R., Przybylski, A. K., Tiokhin, L., Wiernik, B. M., and Orben, A. (2023). Not all effects are indispensable: Psychological science requires verifiable lines of reasoning for whether an effect matters. {\em Perspectives on Psychological Science} {\bf 18}, 503--507.

\bibitem Baguley, T. (2009). Standardized or simple effect size: What should be reported? {\em British Journal of Psychology} {\bf 100}, 603--617.

\bibitem Beall, A. T., and Tracy, J. L. (2013).  Women are more likely to wear red or pink at peak fertility.  {\em Psychological Science} {\bf 24}, 1837--1841.

\bibitem Bryan, C. J., Tipton, E., and Yeager, D. S. (2021). Behavioural science is unlikely to change the world without a heterogeneity revolution. {\em Nature Human Behaviour}, {\bf 5}(8), 980--989.

\bibitem Button, K. S., Ioannidis, J. P., Mokrysz, C., Nosek, B. A., Flint, J., Robinson, E. S., and Munafò, M. R. (2013). Power failure: Why small sample size undermines the reliability of neuroscience. {\em Nature Reviews Neuroscience} {\bf 14}, 365--376.

\bibitem Chang, W., Cheng, J., Allaire, J., Sievert, C., Schloerke, B., Xie, Y., Allen, J., McPherson, J., Dipert, A., and Borges, B. (2024). shiny: Web Application Framework for R. R package version 1.8. 1.9001, \url{https://CRAN.R-project.org/package=shiny}. 

\bibitem Cohen, J. (1962). The statistical power of abnormal social psychological research: A review. {\em Journal of Abnormal Social Psychology} {\bf 65}, 145--153.

%\bibitem Dehejia, R., and Wahba, S. (1999).  Causal effects in non-experimental studies:  Re-evaluating the evaluation of training programs. {\em Journal of the American Statistical Association} {\bf 94}, 1053--1062.

\bibitem Emirbayer, M. (1997). Manifesto for a relational sociology. {\em American Journal of Sociology} {\bf 103}, 281--317.

\bibitem Garthwaite, P. H., Kadane, J. B., and O'Hagan, A. (2005). Statistical methods for eliciting probability distributions. {\em Journal of the American Statistical Association}, {\bf 100}, 680--701.

\bibitem Gelman, A. (2013).  Too good to be true.  {\em Slate}, 24 July.
\url{https://slate.com/technology/2013/07/statistics-and-psychology-multiple-comparisons-give-spurious-results.html}

\bibitem Gelman, A. (2015).   The connection between varying treatment effects and the crisis of unreplicable research:  A Bayesian perspective. {\em Journal of Management} {\bf 41}, 632--643. 

\bibitem Gelman, A. (2018).  The failure of null hypothesis significance testing when studying incremental changes, and what to do about it. {\em Personality and Social Psychology Bulletin} {\bf 44}, 16--23.

\bibitem Gelman, A. (2019). Post-hoc power using observed estimate of effect size is too noisy to be useful. {\em Annals of Surgery} {\bf 270}, e64.

\bibitem Gelman, A., Hullman, J., and Kennedy, L. (2023).  Causal quartets: Different ways to attain the same average treatment effect.  \url{https://arxiv.org/abs/2302.12878}

\bibitem Gelman, A., and Loken, E. (2014).  The statistical crisis in science. {\em American Scientist} {\bf 102}, 460--465. 

\bibitem Gelman, A., and Margalit, Y. (2021).  Social penumbras predict political attitudes.  {\em Proceedings of the National Academy of Sciences} {\bf 118} (6), e2019375118.

\bibitem Gelman, A., and Weakliem, D. (2009). Of beauty, sex and power: Too little attention has been paid to the statistical challenges in estimating small effects. {\em American Scientist} {\bf 97}, 310--316.

\bibitem Gertler, P., Heckman, J., Pinto, R., Zanolini, A., Vermeerch, C., Walker, S., Chang, S. M., and Grantham-McGregor, S. (2013).  Labor market returns to early childhood stimulation intervention in Jamaica.  Institute for Research on Labor and Employment working paper \#142-13.

\bibitem Hill, C. J., Bloom, H. S., Black, A. R.,and Lipsey, M. W. (2008). Empirical benchmarks for interpreting effect sizes in research. {\em Child development perspectives}, {\bf 2}(3), 172--177.

\bibitem Hone, L. S. E., and McCullough, M. E. (2020).  Are women more likely to wear red and pink at peak fertility?  What about on cold days?  Conceptual, close, and extended replications with novel clothing colour measures.  {\em British Journal of Social Psychology} {\bf 59}, 945--964.

\bibitem Hotz, V. J., Imbens, G., and Mortimer, J. (2005). Predicting the efficacy of future training programs using past experiences at other locations.  {\em Journal of Econometrics} {\bf 125}, 241--270.

\bibitem Kennedy, L., and Gelman, A. (2021).  Know your population and know your model: Using model-based regression and poststratification to generalize findings beyond the observed sample.  {\em Psychological Methods} {\bf 26}, 547--558.

\bibitem Kenny, D. A., and Judd, C. M. (2019). The unappreciated heterogeneity of effect sizes: Implications for power, precision, planning of research, and replication. {\em Psychological Methods} {\bf 24}(5), 578.

\bibitem Linden, A. H. (2019). Heterogeneity of research results: New perspectives on psychological science. Doctoral dissertation, Northumbria University.

\bibitem Linden, A. H., and Hönekopp, J. (2021). Heterogeneity of research results: A new perspective from which to assess and promote progress in psychological science. {\em Perspectives on Psychological Science} {\bf 16}, 358--376.

%\bibitem Matejka, J., and Fitzmaurice, G. (2017).  Same stats, different graphs: Generating datasets with varied appearance and identical statistics through simulated annealing. {\em Proceedings of the 2017 CHI Conference on Human Factors in Computing Systems (CHI '17)}, 1290--1294.

\bibitem Mikkola, P., Martin, O. A., Chandramouli, S., Hartmann, M., Pla, O. A., Thomas, O., Pesonen, H., Corander, J., Vehtari, A., Kaski. S., Bürkner, P. C. and Klami, A. (2021). Prior knowledge elicitation: The past, present, and future. {\em Bayesian Analysis} {\bf 19}, 1129--1161.

\bibitem McShane, B. B., Gal, D., Gelman, A., Robert, C., and Tackett, J. L. (2017). Abandon statistical significance. {\em American Statistician} {\bf 73} (S1), 235--245.

\bibitem McShane, B. B., Tackett, J. L., Böckenholt, U., and Gelman, A. (2019). Large-scale replication projects in contemporary psychological research. {\em American Statistician}, {\bf 73}(sup1), 99-105.

\bibitem Mischel, W. (1968).  {\em Personality and Assessment}. New York: Wiley.

\bibitem Myors, B., Murphy, K. R., and Wolach, A. (2014). {\em Statistical Power Analysis: A Simple and General Model for Traditional and Modern Hypothesis Tests}. Routledge.

\bibitem Oakley, J. (2024). {\em SHELF: Tools to Support the Sheffield Elicitation Framework}. R package version 1.10.0, \url{https://github.com/OakleyJ/SHELF}.

\bibitem O'Hagan, A., Buck, C. E., Daneshkhah, A., Eiser, J. R., Garthwaite, P. H., Jenkinson, D. J., Oakley, J., and Rakow, T. (2006). {\em Uncertain Judgements: Eliciting Experts' Probabilities}. Wiley.

\bibitem O'Hagan, A. (2019) Expert knowledge elicitation: Subjective but scientific. {\em American Statistician} {\bf 73}, 69--81.

%\bibitem Popper, K. (1934). {\em The Logic of Scientific Discovery}. Routledge.

\bibitem Rubin, D. B. (1989). A new perspective on meta-analysis. In {\em The Future of Meta-Analysis}, ed.\ K. W. Wachter and M. L. Straff, 155--165.  New York: Russell Sage Foundation.

\bibitem Scargle, J. D. (2000). Publication bias (the ``file-drawer problem'') in scientific inference. {\em Journal of Scientific Exploration} {\bf 14}, 91--106.

\bibitem Simmons, J., Nelson, L., and Simonsohn, U. (2011). False-positive psychology: Undisclosed flexibility in data collection and analysis allow presenting anything as significant. {\em Psychological Science} {\bf 22}, 1359--1366.

\bibitem Singal, A. G., Higgins, P. D., and Waljee, A. K. (2014). A primer on effectiveness and efficacy trials. {\em Clinical and Translational Gastroenterology} {\bf 5}, e45.

\bibitem Stanley, T. D., Carter, E. C., and Doucouliagos, H. (2018). What meta-analyses reveal about the replicability of psychological research. {\em Psychological Bulletin} {\bf 144} (12), 1325--1346.

\bibitem Szászi, B., Higney, A. C., Charlton, A. B., Gelman, A., Ziano, I., Aczel, B., Goldstein, D. G., Yeager,  D. S., and Tipton, E. (2022). No reason to expect large and consistent effects of nudge interventions. {\em Proceedings of the National Academy of Sciences} {\bf 119}, e2200732119.
  
\bibitem Taylor, J. A., Kowalski, S. M., Polanin, J. R., Askinas, K., Stuhlsatz, M. A., Wilson, C. D., Tipton, E., and Wilson, S. J. (2018). Investigating science education effect sizes: Implications for power analyses and programmatic decisions. {\em AERA Open} {\bf 4(3)}.

\bibitem Tipton, E., and Olsen, R. (2018). A review of statistical methods for generalizing from evaluations of educational interventions.  {\em Educational Researcher} {\bf 47}, 516--524. 

\bibitem Tipton, E., Bryan, C., Murray, J., McDaniel, M., Schneider, B., and Yeager, D. S. (2023). Why meta-analyses of growth mindset and other interventions should follow best practices for examining heterogeneity: Commentary on Macnamara and Burgoyne (2023) and Burnette et al.\ (2023). {\em Psychological Bulletin} {\bf 149}, 229--241.

\bibitem Tversky, A., and Kahneman, D. (1974). Judgment under uncertainty: Heuristics and biases. {\em Science} {\bf 185}, 1124--1131.
 
\bibitem van Zwet, E. (2018). A default prior for regression coefficients. {\em Statistical Methods in Medical Research} {\bf 28}, 3799--3807.

\bibitem van Zwet, E., and Gelman, A. (2022). A proposal for informative default priors scaled by the standard error of estimates. {\em American Statistician} {\bf 76}, 1--9.

\bibitem van Zwet, E., Wiecek, W., and Gelman, A. (2025). Meta-analysis with a single study. {\em Statistical Methods in Medical Research}.

\bibitem Wu, Y., Guo, Z., Mamakos, M., Hartline, J., and Hullman, J. (2024). The rational agent benchmark for data visualization. {\em IEEE Transactions on Visualization and Computer Graphics} {\bf 30}, 338--347.

\bibitem Yeager, D. S., Hanselman, P., Walton, G. M., Murray, J. S., Crosnoe, R., Muller, C., \dots and Dweck, C. S. (2019). A national experiment reveals where a growth mindset improves achievement. {\em Nature} {\bf 573}, 364--369.

\bibitem Zelner, J., Riou, J., Etzioni, R., and Gelman, A. (2021).  Accounting for uncertainty during a pandemic. {\em Patterns} {\bf 2}, 100310.

\end{document}